\documentclass[12pt]{article}
\usepackage[pdftex]{graphicx}

\newcommand{\alfven}{Alfv\'{e}n\ }
\newcommand{\alfvenic}{Alfv\'{e}nic\ }
\newcommand{\article}[7]{#1:\ #7,\ {\it #3} {\bf #4},\ #5.}
\newcommand{\articid}[6]{#1:\ #6,\ {\it #3} {\bf #4},\ #5.}
\newcommand{\mib}[1]{\mbox{\boldmath $#1$}}
\newcommand{\SI}    {\Sigma_I}
\newcommand{\SA}    {\Sigma_A}

\newcommand{\asr}{Adv.\ Space Res.}

\newcommand{\apj}{Astrophys. J.}
\newcommand{\eps}{Earth Planets Space}
\newcommand{\grl}{Geophys.\ Res.\ Lett.}
\newcommand{\jgr}{J.\ Geophys.\ Res.}
\newcommand{\mnr}{Mon.\ Not.\ Roy.\ Astron.\ Soc.}
\newcommand{\nat}{Nature}
\newcommand{\pre}{Phys.\ Rev.\ E}
\newcommand{\pss}{Planet.\ Space\ Sci.}
\newcommand{\sci}{Science}
\newcommand{\sop}{Solar Phys.}

\begin{document}

\vspace*{1.0 cm}

\begin{center}

{\large\bf
Self-similar expansion model of cylindrical flux ropes combined with \alfven
 wave current system
}

\vspace{0.5 cm}

Hironori Shimazu\\
National Institute of Information and Communications Technology,
Seika, Kyoto 619-0289, Japan\\
Email: hironori.shimazu@gmail.com\\

\vspace{0.5 cm}

February 13, 2019
\end{center}

\newpage

\begin{center}
{\bf ABSTRACT}
\end{center}

Magnetic flux ropes in space are generally connected to some regions
 electromagnetically.
We consider the whole closed current system of the expanding flux ropes
 including the electric current associated with them.
By combining the theories regarding the self-similar expansion of cylindrical
 flux ropes and the \alfven wave current system, we examine conditions under
 which the electric current matches.
These matching conditions are satisfied when the time dependence of the current
 flowing in the closed circuit agrees with that which maintains the expanding
 flux rope.
In consequence, we encountered three possible forms of expansion.
The two-step eruption of solar filaments may be interpreted as a transition
 from one form of expansion to another.
If this process works, increasing the diffusion outside of the flux rope is
 necessary to trigger the transition.

\newpage

\section{Introduction}

Magnetic flux ropes are magnetohydrodynamics (MHD) structures, in which the
 force-free state of the magnetic field is almost maintained by the electric
 current flowing along the field lines.
They are often observed in eruptive prominences and cores of
coronal mass ejections (CMEs) in the field of view of space-borne coronagraphs
 (Gary and Moore, 2004; Patsourakos et al., 2013; Cheng et al., 2014; Joshi et
 al., 2014).
While propagating in interplanetary space, flux ropes expand with distance from
 the Sun (Burlaga et al., 1981; Marubashi, 1986).
Following the launch of CMEs, the magnetic structure measured in space is
 identified with the flux rope arriving near the Earth (Lepping et al., 1990;
 Dasso et al., 2007).
While some doubts have been raised whether flux ropes exist before eruptions
 (Panasenco et al., 2014), many alternative configurations transform into flux
 ropes via reconnection at the start of the eruptive process (DeVore and
 Antiochos, 2000; Aulanier et al., 2010).

The expansion of the flux rope has often been considered to be self-similar.
The technique of converting time derivatives into self-similar expansion
 parameters in MHD equations was originally applied by Bernstein and Kulsrud
 (1965) and Kulsrud et al. (1965) to supernova explosions.
The self-similar approach simplifies the time-dependent problems and makes them
 analytically tractable.
Low (1982) found a class of spherical self-similar solutions of the expanding
 solar corona.
Kumar and Rust (1996) showed that by assuming the total magnetic helicity is
 conserved a flux rope evolves self-similarly.
Gibson and Low (1998) used the same approach, and presented a theoretical MHD
 model describing the time-dependent expulsion of a CME.

The models of the expansion of the cylindrical flux ropes can be categorized as
 either expansions in all directions,
called 3D expansions,
or those that do not include an axial expansion, i.e., only radial expansion,
called 2D expansions (see Figure 1).
Osherovich et al. (1993; 1995) analyzed the MHD equations and
found a class of self-similar solutions for 2D expansions.
They showed that for this class of self-similar solutions cylindrical flux
 ropes continued to expand only when the polytropic index $\gamma$ is less than
 1.

Models of the 3D self-similar expansions for cylindrical flux ropes were
 developed by Shimazu and Vandas (2002) and Berdichevsky et al. (2003).
Shimazu and Vandas (2002) showed theoretically that flux ropes continue to
 expand self-similarly in medium of any $\gamma$.
That is, 3D expansions are more applicable in space.
Theoretical models (Chen and Garren, 1993; Chen, 1996) and MHD simulations
 (Vandas et al., 1995; Wu et al., 1997; Odstr\v{c}il and Pizzo, 1999; Vandas
 and Odstr\v{c}il, 2000) also showed that flux ropes continued to expand when
 $\gamma$ is larger than 1.

Flux ropes in space are generally connected to some regions electromagnetically
 unless they are closed structures such as tori.
For example, magnetic field lines of a flux rope emerging from the Sun are
 often connected to the solar surface, and must close.
The field-aligned electric current flowing in the flux rope must also close.
Hence, we must consider the whole current system including the electric current
 flowing outside of the flux ropes.

The total electric current flowing in the flux rope decreases for 3D expansions
 (Shimazu and Vandas, 2002).
Nevertheless, because it forms a closed circuit, the change in current flowing
 in the flux rope, which is controlled by the expansion, should affect the
 exterior current in or around the Sun.

The flux rope expansion generates \alfven waves because the deformation of the
 magnetic field lines accompanies the expansion.
The change in the field-aligned current also generates \alfven waves, which
 carry an electric current.
The current accompanied by the \alfven waves compensates the decrease in
 current in the 3D expansion.

In previous studies of flux rope expansions (for example, Shimazu and Vandas
 (2002)), the \alfven wave radiation and the electric current closure outside
 of the flux rope were not considered.
In this paper, the electric current accompanied by the \alfven waves is
 considered.
We examine the current matching conditions by combining the theories underlying
 the self-similar expansion of flux ropes and the \alfven wave current system.

\section{Results}

\subsection{2D self-similar expansion}

Using a cylindrical coordinate system ($r$, $\theta$, $z$) that moves with the
 flux rope, solutions were sought to the MHD equations,
\begin{equation}
\frac{\partial \rho}{\partial t} + \nabla \cdot (\rho \mib{v})=0,
 \label{eq:mhd1}
\end{equation}
\begin{equation}
\rho \frac{\partial \mib{v}}{\partial t} + \rho (\mib{v} \cdot \nabla) \mib{v}
 = - \nabla P + \frac{1}{\mu} ( \nabla \times \mib{B}) \times \mib{B},
 \label{eq:mhd2}
\end{equation}
\begin{equation}
\frac{\partial (P {\rho}^{-\gamma})}{\partial t} + (\mib{v} \cdot \nabla )(P
 {\rho}^{-\gamma})=0, \label{eq:mhd3}
\end{equation}
and
\begin{equation}
\frac{\partial \mib{B}}{\partial t} = \nabla \times (\mib{v} \times \mib{B}),
 \label{eq:mhd4}
\end{equation}
where $t$ is the elapsed time, $\rho$ the mass density, \mib{v} the velocity,
 $P$ the pressure, $\mu$ permeability, \mib{B} the magnetic field, and $\gamma$
 the polytropic index.
The $z$-axis is taken to lie along the axis of the cylindrical flux rope;
the solution is assumed to have no dependence on $z$ ($r$-dependence only) for
 2D expansions.

Following the procedure described in Osherovich et al. (1995),
we let $f$ denote the generating function of the self-similar parameter $\eta$,
 and $y$ the evolution function of time.
The solutions of Equations (\ref{eq:mhd1}), (\ref{eq:mhd3}), and
 (\ref{eq:mhd4}) for 2D expansions are then
\begin{equation}
v_r = \eta \dot{y},
\end{equation}
\begin{equation}
B_{\theta} = (-\eta f' /2)^{1/2} y^{-1}, \label{eq:bto}
\end{equation}
\begin{equation}
B_{z} = (2 \mu S D)^{1/2} y^{-2}, \label{eq:bzo}
\end{equation}
\begin{equation}
\rho = -D' \eta ^{-1} y^{-2},
\end{equation}
and
\begin{equation}
P = KD y^{-2 \gamma},
\end{equation}
where an overdot signifies the derivative with respect to time $t$, and a prime
 signifies the derivative with respect to $\eta$, which satisfies
\begin{equation}
\eta = ry^{-1},
\end{equation}
\begin{equation}
D = \frac{f+\eta f' /2}{2 \mu S \chi}, \label{eq:ddef}
\end{equation}
and $\chi$, $S$, and $K$ are positive constants.
The function $f$ satisfies
\begin{equation}
f' \leq 0
\end{equation}
\begin{equation}
f + \eta f' /2 \geq 0
\end{equation}
and
\begin{equation}
(f + \eta f' /2)' \leq 0.
\end{equation}

We take $K \rightarrow 0$ (low plasma beta value).
Moreover, being most frequently used in the description of a cylindrical flux
 rope, $f$ is taken to be in the form
\begin{equation}
f=B_1^2 \{ J_0^2(\alpha _1 \eta )+J_1^2(\alpha _1 \eta ) \}, \label{eq:gen}
\end{equation}
where $J_0$ and $J_1$ are Bessel functions of the first kind of orders 0 and 1,
 respectively, and $B_1$ and $\alpha _1$ are constants.
Hence, the magnetic field of the 2D expansion model is given by
\begin{equation}
B_{\theta} = B_0J_1(\alpha ry ^{-1} ) y^{-1}, \label{eq:woe}
\end{equation}
\begin{equation}
B_{z} = B_0J_0(\alpha r y^{-1} ) y^{-2}, \label{eq:woez}
\end{equation}
where
\begin{equation}
y = 1+t/t_0,
\end{equation}
and $B_0$, $t_0$, and $\alpha$ are constants.

\subsection{3D self-similar expansion}

We include the effects of an axial expansion ($z$-direction) as well as a
 radial expansion.
An additional self-similar parameter $\xi$ is introduced by
\begin{equation}
\xi = zy^{-1}.
\end{equation}
We assume that the radial expansion rate is the same as the axial expansion
 rate.
The solutions of Equations (\ref{eq:mhd1}), (\ref{eq:mhd3}), and
 (\ref{eq:mhd4}), which satisfy
\begin{equation}
\frac{\partial f}{\partial \xi}=0   \label{eq:zd}
\end{equation}
in the simplest case are given by
\begin{equation}
v_r = \eta \dot{y},
\end{equation}
\begin{equation}
v_z = \xi \dot{y},
\end{equation}
\begin{equation}
B_{\theta} = (-\eta f' /2)^{1/2} y^{-2}, \label{eq:btn}
\end{equation}
\begin{equation}
B_{z} = (2 \mu S D)^{1/2} y^{-2}, \label{eq:bzn}
\end{equation}
\begin{equation}
\rho = -G' \eta ^{-1} y^{-3},
\end{equation}
and
\begin{equation}
P = KG y^{-3 \gamma},
\end{equation}
where $G$ is a function of $\eta$ and $\xi$ (Shimazu and Vandas, 2002).

With $K \rightarrow 0$ and $f$ as given in Equation (\ref{eq:gen}),
the magnetic field of the flux rope for the 3D expansion is expressed as
\begin{equation}
B_{\theta} = B_0J_1(\alpha r y^{-1}) y^{-2}, \label{eq:el}
\end{equation}
\begin{equation}
B_{z} = B_0J_0(\alpha r y^{-1}) y^{-2}. \label{eq:ez}
\end{equation}

The expression for $B_z$ is similar to the field solution in the 2D expansion
 model.
The difference in the dependence of $B_{\theta}$ on $y$ (or $t$) stems from the
 volume increase in the axial direction.
This is the essential effect of the axial expansion.
As is easily demonstrated, the total electric current is conserved in the 2D
 expansion model,
whereas the magnetic helicity and the magnetic flux are not conserved.
In contrast, the magnetic flux and magnetic helicity in the flux rope is
 conserved in the 3D expansion model,
whereas the total electric current decreases with time (Shimazu and Vandas,
 2002).

\subsection{\alfven wave current system}

We adopt \alfven wing theory for the electric current system generated by the
 \alfven waves.
This theory was developed originally to describe the interaction between a
 flowing magnetized plasma and a conductor (Drell et al., 1965; Wright and
 Schwartz, 1990).
Consider a conductor moving across a uniform magnetic field \mib{B} in a plasma
 with a velocity \mib{v}.
The \alfven waves radiate from the polarization charges and carry an electric
 current into the surrounding plasma.
The magnetic field lines in the plasma act as transmission lines for \alfven
 waves.
The region through which the \alfven waves propagate is called the ``\alfven
 wing".
The theory of the \alfven wing was advanced largely in studies of the
 electromagnetic coupling between the Jovian magnetosphere and the Jovian
 satellite, Io (Neubauer, 1980; Goertz, 1980; Southwood et al., 1980).
The Voyager and Galileo satellites detected disturbances of the plasma velocity
 and magnetic fields associated with the generation of the \alfven wing around
 Io (Acu\~{n}a et al., 1981; Chust et al., 2005).

From the current continuity condition, the induced current $J$ flowing in the
 conductor and the \alfven wing is expressed as
\begin{equation}
J = |\mib{v} \times \mib{B}| a \frac{2 \SI \SA}{\SI +2\SA} , \label{eq:ji}
\end{equation}
where $a$ is the length scale of the voltage or current generator,
$\SI$ the conductance of the conductor, and
$\SA$ the \alfven conductance, which is related to the polarization current
 flowing in the wave front (Hill et al., 1983).
If $v/V_A \ll 1$, we have
\begin{equation}
\SA \equiv \frac{1}{\mu V_A }, \label{eq:alc}
\end{equation}
where $V_A \equiv B /\sqrt{\mu \rho}$ is the \alfven velocity (Neubauer, 1980).
Note that $\SA$ is finite, even though the plasma conductivity is infinitely
 large.
In addition, $J$ remains finite, even as $\SI$ tends to infinity; see Equation
 (\ref{eq:ji}).
We consider two extreme cases (Shimazu and Terasawa, 1995):
$\SI \gg \SA$,
called an \alfvenic case with
\begin{equation}
J = 2 |\mib{v} \times \mib{B}| a \SA , \label{eq:ondo}
\end{equation}
and $\SI \ll \SA$
a diffusive case with
\begin{equation}
J = |\mib{v} \times \mib{B}| a \SI . \label{eq:low}
\end{equation}

\subsection{Combining the two theories}

In this paper, we include the electric current associated with the flux rope
 and consider the whole closed current system.
We exclude isolated flux ropes that are not connected elsewhere
 electromagnetically.
Consider a flux rope expanding near the solar surface
and divide the whole closed current circuit into two regions (Figure 2):
the expanding flux rope and the connected region (region excluding the flux
 rope).
The expansion of the flux rope generates \alfven waves through the deformation
 of the magnetic field lines and
the change in the field-aligned current.
The current flowing in the flux rope must close with that carried by the
 \alfven waves.

This system is equivalent to an electric circuit (Figure 3)
with $\SI$ representing the conductance of the connected region, and $\SA$ the
 \alfven conductance in the flux rope.
The expansion of the flux rope corresponds to the source of current or voltage.
They constitute a closed circuit.

We apply the \alfven wing theory to the \alfven wave current system generated
 by the flux rope expansion.
Of course, an \alfven wing is not generated in the system we are considering.
However, a similar current circuit as in the \alfven wing system is generated
 in this system.
To apply the \alfven wing theory, we examine the dependence of $J$ on $y$ by
 combining the self-similar flux rope expansion model.

In the expanding flux rope, the electric field in the axial direction is
 generated by the radial expansion.
The electric field is given by
$| - \mib{v} \times \mib{B} |_z = |v_r B_ \theta | \sim y^{-2}$ for the 3D
 expansion and $\sim y^{-1}$ for the 2D expansion, respectively.
As $\SA$ has the form $\sim \rho^{1/2} B_ \theta ^{-1}$,
$
\SA
\sim y^{1/2}
$
for 3D expansions, and
$
\SA
\sim y^{0}
$
for 2D expansions (no dependence on $y$).
Therefore, in the \alfvenic case, Equation (\ref{eq:ondo}) leads to
\begin{equation}
J
\sim y^{-1/2}, \label{eq:3db}
\end{equation}
for 3D expansions,
and
\begin{equation}
J
\sim y^{0}, \label{eq:2db}
\end{equation}
for 2D expansions.
Note that $a \sim v_r y$.
When $\SI$ is assumed constant,
Equation (\ref{eq:low}) in the diffusive case leads to
\begin{equation}
J
\sim y^{-1}, \label{eq:3dc}
\end{equation}
for 3D expansions and
\begin{equation}
J
\sim y^{0}, \label{eq:2dc}
\end{equation}
for 2D expansions.

Because $J$ must agree with the current maintaining the magnetic field of the
 flux rope,
\begin{equation}
J
 = \frac{2 \pi}{\mu} \int ^a _0 \frac{\partial B_ \theta}{\partial r} rdr,
\end{equation}
the dependence of $J$ on $y$ is given by
\begin{equation}
J \sim y^{-1}, \label{eq:3df}
\end{equation}
for 3D expansions, and
\begin{equation}
J \sim y^{0}, \label{eq:2df}
\end{equation}
for 2D expansions.

We compare Equations (\ref{eq:3df}) and (\ref{eq:2df}) with Equations
 (\ref{eq:3db}) -- (\ref{eq:2dc}).
For 3D diffusive cases, the dependences on $y$ in Equations (\ref{eq:3dc}) and
 (\ref{eq:3df}) agree;
for 2D diffusive cases, Equations (\ref{eq:2dc}) and (\ref{eq:2df}) agree.
These agreements show that closure is attained between the current flowing in
 the flux rope and the real current flowing in the connected region.
Hence, diffusion ($\SI$) is necessary in the connected region.
In this diffusive expansion, both 3D and 2D expansions are possible.
However, the 3D diffusive expansion is more applicable because the expansion
 continues when $\gamma$ is larger than 1.

In addition, dependences on $y$ in Equations (\ref{eq:2db}) and (\ref{eq:2df})
 also agree for the 2D \alfvenic case.
In this case, the current flowing in the flux rope closes with that carried by
 the \alfven waves.
Table 1 shows the possible combinations for these expansions.

\section{Discussion}

To summarize the previous section, the 3D diffusive expansion and the 2D
 diffusive expansion are possible when $\SI \ll \SA$.
A possible alternative case is the 2D \alfvenic expansion when $\SI \gg \SA$.

Two-step solar filament eruptions may be interpreted as a transition from one
 expansion to another.
Sometimes the filament eruption decelerates and stops at some height in the
 corona.
After several hours the filament rises again and forms a CME (Byrne et al.,
 2014; Gosain et al., 2016; Chanda et al., 2017).
One of the attractive models of filament eruptions is the catastrophes in the
 system equilibrium.
Van Tend and Kuperus (1978) showed that the equilibrium of a linear electric
 current in the coronal magnetic field is unstable depending on spatial
 properties of the coronal field.
Priest and Forbes (1990) analyzed in detail the equilibrium and dynamics of a
 straight flux tube in a background magnetic field of a horizontal dipole
 located below the conductive surface (photosphere).
They showed that a loss of equilibrium in the system causes an eruption of the
 filament (Schmieder et al., 2015; Filippov, 2018).

Here, we apply our flux rope expansion model to the two-step filament
 eruptions.
When the flux rope starts to expand, \alfven waves are generated carrying an
 electric current.
The expansion is Alfv\'{e}nic because the Lundquist number $\SI / \SA = \mu V_A
 \SI$ is much larger than 1.
Even in weakly ionized regions such as the photosphere and the chromosphere,
 the Lundquist number is larger than $10^4$.
In \alfvenic expansions, only the 2D expansion is possible, and
because it does not involve an axial expansion,
the ascension of the flux rope is relatively small.
Moreover, the 2D expansion does not continue when $\gamma$ is larger than 1,
 and hence the expansion stops at some level.

After the launch of a CME, the flux rope propagates in interplanetary space.
As $\SI / \SA \ll 1$, diffusive expansion occurs.
Note that $\SI$ is for the connected region in or very close to the Sun and
 $\SA$ is for the flux rope in interplanetary space.
Because an axial expansion accompanies a 3D expansion,
the flux rope ascends rapidly.
To trigger the transition from a 2D \alfvenic expansion to a 3D diffusive
 expansion, there should be some processes to reduce $\SI$ in the connected
 region.
For example, reconnection in the connected region and a change in the current
 circuit may trigger a transition.

If this process works, the initial 2D \alfvenic expansion is a source of
 \alfven waves in the solar atmosphere.
There is much research showing that \alfven waves are responsible for heating
 of the solar atmosphere (Osterbrock, 1961; Ionson, 1978; Hollweg, 1991;
 Sakurai et al., 1991).
\alfven or transverse mode MHD waves have been observed in the solar
 photosphere, chromosphere, and corona in some detail using instruments onboard
 satellite missions and ground-based solar telescopes (Aschwanden et al., 1999;
 Nakariakov et al., 1999; Ofman and Wang 2008; McIntosh et al., 2011; Okamoto
 et al., 2015; Antolin et al., 2015).

\section{Summary}

We included the electric current associated with the expanding magnetic flux
 rope and considered the whole closed current system.
By combining the theories of the self-similar expansion of cylindrical flux
 ropes and the \alfven wave current system,
we examined those conditions for which
the time dependence of the current flowing in the closed circuit agrees with
 that of the current sustaining the expanding flux rope.

The results have shown that there are three possible expansions: 3D diffusive,
 2D diffusive, and 2D \alfvenic expansions.
The 3D and 2D diffusive expansions occur when $\SI / \SA \ll 1$.
In this case, diffusion in the connected region is necessary for the flux rope
 expansion.
The remaining case is the 2D \alfvenic expansion occurring when $\SI / \SA \gg
 1$.
The current flowing in the flux rope then closes with the current carried by
 the \alfven waves.

The two-step solar filament eruptions may be interpreted as a transition from
 the 2D \alfvenic expansion to the 3D diffusive expansion.
When the flux rope starts to expand, \alfven waves are generated and carry the
 electric current.
The 2D \alfvenic expansion occurs, because $\SI / \SA \gg 1$.
Because the 2D expansion does not involve an axial expansion,
the ascension of the flux rope is relatively small.
In addition, the expansion does not continue, because $\gamma$ is larger than 1
 and thus, stops at some level.

After the launch of a CME, the flux rope propagates in interplanetary space.
Because $\SI / \SA \ll 1$, a 3D diffusive expansion should occur.
Note that $\SI$ is for the connected region in or very close to the Sun and
 $\SA$ is for the flux rope in interplanetary space.
Because an axial expansion accompanies the 3D expansion,
the flux rope ascends rapidly.
To trigger the transition from a 2D \alfvenic expansion to a 3D diffusive
 expansion, there should be some processes to increase the diffusion in the
 connected region.

Our interpretation of the two-step filament eruptions did not include
 considering the stability of the flux rope in the background plasma and
 magnetic field.
We also did not include the curvature of the flux rope axis, the details of the
 electric current closure or the \alfven wave propagation outside of the flux
 rope.
We had restricted our attention solely to self-similar expansions.
However, this study is our first step with the new interpretation.
We intend next to study its consistency with respect to the theory of the loss
 of equilibrium.

\vspace{1cm}

\begin{flushleft}
{\bf Acknowledgments}
\end{flushleft}
The author thanks
Yasuhiro Murayama (National Institute of Information and Communications
 Technology, Japan)
for his stimulating and insightful comments and suggestions during the course
 of this work.
This work was supported by NICT 320163 and 190102.

\newpage

\begin{flushleft}
{\bf References}
\end{flushleft}

\begin{myreference}

\article
{Acu\~{n}a, M.H., Neubauer, F.M., Ness, N.F.}
{Standing \alfven wave current system at Io: Voyager 1 observations}
{\jgr}{86}{8513}{8521}{1981}

\articid
{Antolin, P., Okamoto, T.J., De Pontieu, B., Uitenbroek, H., Van Doorsselaere,
 T., Yokoyama, T.}
{Resonant Absorption of Transverse Oscillations and Associated Heating in a
 Solar Prominence. II. Numerical Aspects}
{\apj}{809}{72}{2015}

\article
{Aschwanden, M.J., Fletcher, L., Schrijver, C.J., Alexander, D.}
{Coronal loop oscillations observed with the transition region and coronal
 explorer}
{\apj}{520}{880}{894}{1999}

\article
{Aulanier G., T\"{o}r\"{o}k T., D\'{e}moulin P., DeLuca E.E.}
{Formation of torus-unstable flux ropes and electric
currents in erupting sigmoids}
{\apj}{708}{314}{333}{2010}

\articid
{Berdichevsky, D.B., Lepping, R.P., Farrugia, C. J.}
{On geometric considerations of the evolution of magnetic flux ropes}
{\pre}{67}{036405}{2003}

\article
{Bernstein, I.B., Kulsrud, R.M.}
{On the explosion of a supernova into the interstellar magnetic field. I}
{\apj}{142}{479}{490}{1965}

\article
{Burlaga, L., Sittler, E., Mariani, F., Schwenn, R.}
{Magnetic loop behind an interplanetary shock: Voyager, Helios, and IMP 8
 observations}
{\jgr}{86}{6673}{6684}{1981}

\article
{Byrne J.P., Morgan H., Seaton D.B., Bain H.M., Habbal S.R.}
{Bridging EUV and white-light observations to inspect the initiation phase of a
 ``Two-Stage" solar eruptive event}
{\sop}{289}{4545}{4562}{2014}

\articid
{Chandra, R., Filippov, B., Joshi, R., Schmieder, B.}
{Two-step filament eruption during 14-15 March 2015}
{\sop}{292}{81}{2017}

\article
{Chen, J.}
{Theory of prominence eruption and propagation: Interplanetary consequences}
{\jgr}{101}{27499}{27519}{1996}

\article
{Chen, J., Garren, D.A.}
{Interplanetary magnetic clouds: Topology and driving mechanism}
{\grl}{20}{2319}{2322}{1993}

\articid
{Cheng, X., Ding, M.D., Guo, Y., Zhang, J., Vourlidas, A., Liu, Y.D., Olmedo,
 O., Sun, J.Q., Li, C.}
{Tracking the Evolution of a Coherent Magnetic Flux Rope Continuously from the
 Inner to the Outer Corona}
{\apj}{780}{28}{2014}

\article
{Chust, T., Roux, A., Kurth, W.S., Gurnett, D.A., Kivelson, M.G., Khurana,
 K.K.}
{Are Io's Alfven wings filamented? Galileo observations}
{\pss}{53}{395}{412}{2005}

\article
{Dasso, S., Nakwacki, M.S., D\'{e}moulin, P., Mandrini, C.H.}
{Progressive Transformation of a Flux Rope to an ICME. Comparative Analysis
 Using the Direct and Fitted Expansion Methods}
{\sop}{244}{115}{137}{2007}

\article
{DeVore, C.R., Antiochos, S.K.}
{Dynamical Formation and Stability of Helical Prominence Magnetic Fields}
{\apj}{539}{954}{963}{2000}

\article
{Drell, S.D., Foley, H.M., Ruderman, M.A.}
{Drag and propulsion of large satellites in the ionosphere:
An \alfven propulsion engine in space}
{\jgr}{70}{3131}{3145}{1965}

\article
{Filippov, B.}
{Two-step solar filament eruptions}
{\mnr}{475}{1646}{1652}{2018}

\article
{Gary, G.A., Moore, R.L.}
{Eruption of a multiple-turn helical magnetic flux tube in a large flare:
 evidence for external and internal reconnection that fits the breakout model
 of solar magnetic eruptions}
{\apj}{611}{545}{556}{2004}

\article
{Gibson, S.E., Low, B.C.}
{A time-dependent three-dimensional magnetohydrodynamic model of the coronal
 mass ejection}
{\apj}{493}{460}{473}{1998}

\article
{Goertz, C.K.}
{Io's interaction with the plasma torus}
{\jgr}{85}{2949}{2956}{1980}

\articid
{Gosain, S., Filippov, B., Ajor Maurya, R., Chandra, R.}
{Interrupted eruption of large quiescent filament associated with a halo CME}
{\apj}{821}{85}{2016}

Hill, T.W., Dessler, A.J., Goertz, C.K.: 1983,
Magnetospheric models,
in {\it Physics of the Jovian Magnetosphere}, edited by A.J. Dessler,
pp. 365-372, Cambridge Univ. Press, New York.

Hollweg, J.V.; 1991, in {\it Mechanisms of Chromospheric and Coronal Heating},
 edited by P. Ulmschneider, E.R. Priest, R. Rosner, p. 423, Springer, Berlin.

\article
{Ionson, J.A.}
{Resonant absorption of Alfvenic surface waves and the heating of solar coronal
 loops}
{\apj}{226}{650}{673}{1978}

\articid
{Joshi, N.C., Srivastava, A.K., Filippov, B., Kayshap, P., Uddin, W., Chandra,
 R., Prasad Choudhary, D., Dwivedi, B.N.}
{Confined partial filament eruption and its reformation within a stable
 magnetic flux rope}
{\apj}{787}{11}{2014}

\article
{Kulsrud, R.M., Bernstein, I.B., Kruskal, M., Fanucci, J., Ness, N.}
{On the explosion of a supernova into the interstellar magnetic field. II}
{\apj}{142}{491}{506}{1965}

\article
{Kumar, A., Rust, D.M.}
{Interplanetary magnetic clouds, helicity conservation, and current-core
 flux-ropes}
{\jgr}{101}{15667}{15684}{1996}

\article
{Lepping, R.P., Jones, J.A., Burlaga, L.F.}
{Magnetic field structure of interplanetary magnetic clouds at 1 AU}
{\jgr}{95}{11957}{11965}{1990}

\article
{Low, B.C.}
{Self-similar magnetohydrodynamics. I. The $\gamma =4/3$ polytrope and the
 coronal transient}
{\apj}{254}{796}{805}{1982}

\article
{Marubashi, K.}
{Structure of the interplanetary magnetic clouds and their solar origins}
{\asr}{6(6)}{335}{338}{1986}

\article
{McIntosh, S.W., de Pontieu, B., Carlsson, M., Hansteen, V., Boerner, P.,
 Goossens, M.}
{\alfvenic waves with sufficient energy to power the quiet solar corona and
 fast solar wind}
{\nat}{475}{477}{480}{2011}

\article
{Nakariakov, V.M., Ofman, L., Deluca, E.E., Roberts, B., Davila, J.M.}
{TRACE observation of damped coronal loop oscillations: Implications for
 coronal heating}
{\sci}{285}{862}{864}{1999}

\article
{Neubauer, F.M.}
{Nonlinear standing \alfven wave current system at Io: Theory}
{\jgr}{85}{1171}{1178}{1980}

\article
{Odstr\v{c}il, D., Pizzo, V.J.}
{Three-dimensional propagation of coronal mass ejections (CMEs) in a structured
 solar wind flow 1. CME launched within the streamer belt}
{\jgr}{104}{483}{492}{1999}

\article
{Ofman, L., Wang, T.J.}
{Hinode observations of transverse waves with flows in coronal loops}
{\apj}{482}{L9}{L12}{2008}

\articid
{Okamoto, T.J., Antolin, P., De Pontieu, B., Uitenbroek, H., Van Doorsselaere,
 T., Yokoyama, T.}
{Resonant absorption of transverse oscillations and associated heating in a
 solar prominence. I. Observational aspects}
{\apj}{809}{71}{2015}

\article
{Osherovich, V.A., Farrugia, C.J., Burlaga, L.F.}
{Nonlinear evolution of magnetic flux ropes 1. Low-beta limit}
{\jgr}{98}{13225}{13231}{1993}

\article
{Osherovich, V.A., Farrugia, C.J., Burlaga, L.F.}
{Nonlinear evolution of magnetic flux ropes 2. Finite beta plasma}
{\jgr}{100}{12307}{12318}{1995}

\article
{Osterbrock, D.E.}
{The heating of the solar chromosphere, plages, and corona by
 magnetohydrodynamic waves}
{\apj}{134}{347}{388}{1961}

\article
{Panasenco O., Martin S.F., Velli M.}
{Apparent solar tornado-like prominences}
{\sop}{289}{603}{622}{2014}

\articid
{Patsourakos S., Vourlidas A., Stenborg G.}
{Direct evidence for a fast coronal mass ejection driven by the prior formation
 and subsequent destabilization of a magnetic flux rope}
{\apj}{764}{125}{2013}

\article
{Priest E.R., Forbes T.G.}
{Magnetic field evolution during prominence eruptions and two-ribbon flares}
{\sop}{126}{319}{350}{1990}

\article
{Sakurai, T., Goossens, M., Hollweg, J.V.}
{Resonant behaviour of MHD waves on magnetic flux tubes. I - Connection
 formulae at the resonant surfaces}
{\sop}{133}{227}{245}{1991}

\article
{Schmieder, B., Aulanier, G., Vr\v{s}nak, B.}
{Flare-CME models: An observational perspective (invited review)}
{\sop}{290}{3457}{3486}{2015}

\article
{Shimazu, H., Terasawa, T.}
{Electromagnetic induction heating of meteorite parent bodies by the primordial
 solar wind}
{\jgr}{100}{16923}{16930}{1995}

\article
{Shimazu, H., Vandas, M.}
{A self-similar solution of expanding cylindrical flux ropes for any polytropic
 index value}
{\eps}{54}{783}{790}{2002}

\article
{Southwood, D.J., Kivelson, M.G., Walker, R.J., Slavin, J.A.}
{Io and its plasma environment}
{\jgr}{85}{5959}{5968}{1980}

\article
{Vandas, M., Fischer, S., Dryer, M., Smith, Z., Detman, T.}
{Simulation of magnetic cloud propagation in the inner heliosphere in
 two-dimensions 1. A loop perpendicular to the ecliptic plane}
{\jgr}{100}{12285}{12292}{1995}

\article
{Vandas, M., Odstr\v{c}il, D.}
{Magnetic cloud evolution: A comparison of analytical and numerical solutions}
{\jgr}{105}{12605}{12616}{2000}

\article
{Van Tend, W., Kuperus, M.}
{The development of coronal electric current systems in active regions and
 their relation to filaments and flares}
{\sop}{59}{115}{127}{1978}

\article
{Wright, A.N., Schwartz, S.J.}
{The equilibrium of a conducting body embedded in a flowing plasma}
{\jgr}{95}{4027}{4038}{1990}

\article
{Wu, S.T., Guo, W.P., Dryer, M.}
{Dynamical evolution of a coronal streamer -- flux rope system II. A
 self-consistent non-planar magnetohydrodynamic simulation}
{\sop}{170}{265}{282}{1997}

\end{myreference}

\clearpage

\begin{table}

\caption{Possible combinations of the expansion and the dependence of $J$ on
 $y$.}

\vspace{1cm}

\begin{center}
  \begin{tabular}{cccc} \hline
\multicolumn{2}{c}{diffusive} & \multicolumn{2}{c}{\alfvenic} \\
\multicolumn{2}{c}{($\SI / \SA \ll 1$)} & \multicolumn{2}{c}{($\SI /\SA \gg
 1$)} \\ \hline
3D & 2D & 3D & 2D \\ \hline
$y^{-1}$ & $y^0$ & - & $y^0$ \\ \hline
  \end{tabular}
\end{center}

\end{table}

\clearpage
\includegraphics[width=15cm]{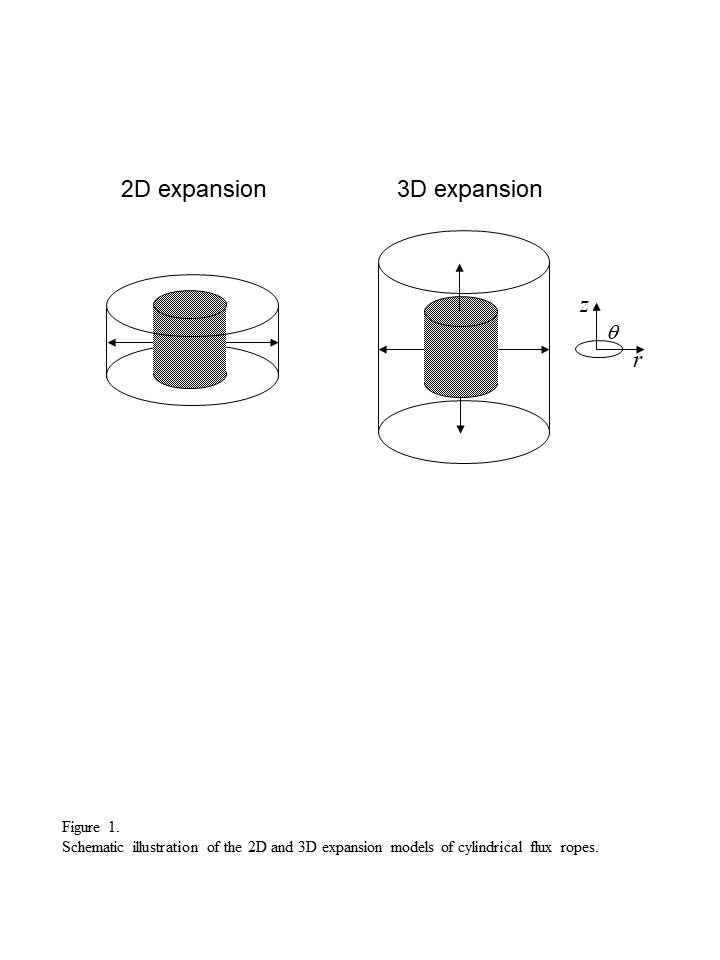}
\newpage
\includegraphics[width=15cm]{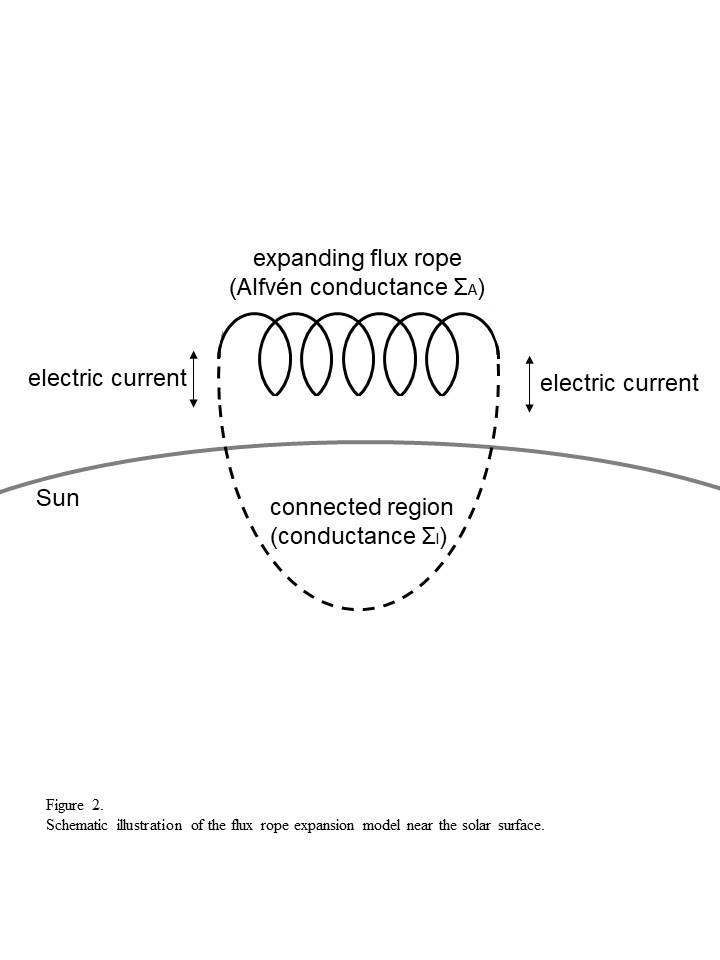}
\newpage
\includegraphics[width=15cm]{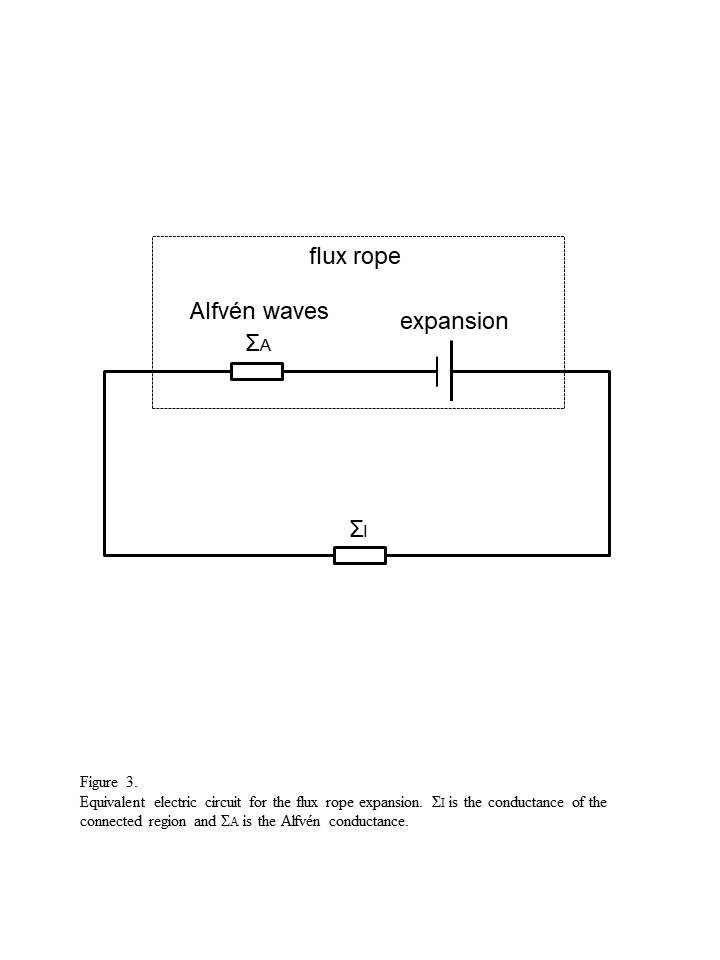}

\end{document}